# A method for real-time volumetric imaging in radiotherapy using single x-ray projection

**Yuan Xu[1, 2], Hao Yan[1], Luo Ouyang[1], Jing Wang[1], Linghong Zhou[2], Laura Cervino[3], Steve B. Jiang[1], and Xun Jia[1]**

[1]Department of Radiation Oncology, University of Texas Southwestern Medical Center, Dallas, TX 75235, USA
[2]Department of Biomedical Engineering, Southern Medical University, Guangzhou, Guangdong 510515, China
[3]Center for Advanced Radiotherapy Technologies and Department of Radiation Medicine and Applied Sciences, University of California San Diego, La Jolla, CA 92037, USA

E-mails: xun.jia@utsouthwestern.edu, steve.jiang@utsouthwestern.edu, smart@smu.edu.cn

It is an intriguing problem to generate an instantaneous volumetric image based on the corresponding x-ray projection. In this paper, we present a new method to achieve this goal via a sparse learning approach. To fully extract motion information hidden in projection images, we partitioned a projection image into small patches. We utilized a sparse learning method to automatically select patches that have a high correlation with principal component analysis (PCA) coefficients of a lung motion model. A model that maps the patch intensity to the PCA coefficients is built along with the patch selection process. Based on this model, a measured projection can be used to predict the PCA coefficients, which are further used to generate a motion vector field and hence a volumetric image. We have also proposed an intensity baseline correction method based on the partitioned projection, where the first and the second moments of pixel intensities at a patch in a simulated projection image are matched with those in a measured one via a linear transformation. The proposed method has been valid in simulated data and real phantom data. The algorithm is able to identify patches that contain relevant motion information, e.g. diaphragm region. It is found that intensity baseline correction step is important to remove the systematic error in the motion prediction. For the simulation case, the sparse learning model reduced prediction error for the first PCA coefficient to 5%, compared to the 10% error when sparse learning is not used, and 95$^{th}$ percentile error for the predicted motion vector is reduced from 2.40 mm to 0.92mm. In the phantom case, the predicted tumor motion trajectory is successfully reconstructed with 0.82 mm mean vector field error compared to 1.66 mm error without using the sparse learning method. The algorithm robustness with respect to sparse level, patch size, and existence of diaphragm, as well as computation time, has also been studied.

## 1. Introduction

Time-resolved volumetric imaging is a key component for motion management in lung radiotherapy (Jiang, 2006; Keall *et al.*, 2006). A number of solutions have been proposed to provide high-quality images at different stages of a radiotherapy treatment, such as 4D-CT for planning (Pan *et al.*, 2004; Keall *et al.*, 2004; Pan *et al.*, 2007; Li *et al.*, 2009), 4D or motion-compensated cone beam CT (CBCT) for pre-treatment setup (Li *et al.*, 2007; Chen *et al.*, 2008; Leng *et al.*, 2008; Rit *et al.*, 2009; Ahmad *et al.*, 2011; Park *et al.*, 2011; Zheng *et al.*, 2011; Jia *et al.*, 2012a; Ren *et al.*, 2012; Gao *et al.*, 2012; Cai *et al.*, 2012; Wang and Gu, 2013; Yan *et al.*, 2014), and various online imaging methods during treatment (Zeng *et al.*, 2005; Low *et al.*, 2005; Zhang *et al.*, 2007; Li *et al.*, 2011a; Li *et al.*, 2011b; Li *et al.*, 2011c; Chou *et al.*, 2011; Staub *et al.*, 2011; Chou *et al.*, 2013; Zhang *et al.*, 2013). In particular, real-time imaging, i.e. reconstructing the instantaneous volumetric patient image, have recently attracted a lot of research attentions due to the great desires of visualizing patient anatomy during a treatment for many clinical tasks, such as tumor tracking, treatment monitoring, and delivered dose assessments.

A volumetric image may be derived based on a variety of different measurements. It is of particular research interest to study single x-ray projection based volumetric image reconstruction, because of the wide availability of CBCT system integrated in a medical linear accelerator (McBain *et al.*, 2006; Jaffray and Siewerdsen, 2000; Jaffray *et al.*, 2002). However, it is not easy to achieve this goal. At a given moment, one can only measure one 2D projection of the 3D volumetric image, and the voxel values along the x-ray projection direction are integrated. Additional information must be supplemented to retrieve each voxel value. One type of prior information is patient-specific lung motion model (Zhang *et al.*, 2007; Staub *et al.*, 2011; Chou *et al.*, 2011; McClelland *et al.*, 2012; Chou *et al.*, 2013). For instance, principal component analysis (PCA) based lung motion model has been proposed. Principal components of deformation vector fields (DVFs) for lung motion are identified and the first a few of them are found to be sufficient to represent the entire lung motion, which greatly reduces degrees of freedom of the problem (Söhn *et al.*, 2005; Zhang *et al.*, 2007; Li *et al.*, 2011b; Li *et al.*, 2011c). This concise representation of lung motion helps the development of real-time volumetric imaging techniques, as the limited amount of measurements at a given moment are now sufficient to determine the PCA coefficients. Based on this idea, we have previously proposed a method and for the first time demonstrated the feasibility to reconstruct the instantaneous volumetric image based on single x-ray projection measurement (Li *et al.*, 2011b). In this method, PCA coefficients of the lung model are adjusted through an optimization process, so that the forward calculated projection of the volumetric image, corresponding to the PCA coefficients, matches the measurements Similar idea has also been employed in 4D-CBCT reconstruction where multiple projections corresponding to a given breathing phase are used (Staub *et al.*, 2011).

Despite the success, there are two major issues associated with this optimization-based approach for real-time imaging. First, the forward matching process tries to minimize intensity difference between the calculated and the measured projections. It is not robust with respect to intensity mismatch between the two due to, for instance, the neglecting of x-ray scatter in the forward calculation. The method works for some patients, but its robustness in general needs to be improved. Second, for a measured x-ray projection, a non-linear optimization problem is solved via a certain iterative algorithm. It is very computationally challenging to achieve a high performance, so that the volumetric image can be updated in a high frequency to be considered real-time.

In this paper, we propose a method to solve the two aforementioned problems. As opposed to going through an optimization-based approach, our method realizes the goal via a mapping-based approach. A relationship between the x-ray projections and the corresponding volume is first established in a learning process. The learned mapping is applied to a measured projection, yielding the volumetric image. Because a simple forward calculation is needed to generate the image based on the projection, the computation efficiency is ensured. We have also developed intensity correction methods and incorporated sparse learning techniques to achieve system robustness.

## 2. Methods and Materials

Generally speaking, there exists a forward mapping between a 3D patient anatomy and an x-ray projection $p = P[f]$, where $p$ is the projection image and $f$ is the 3D volumetric data. Note this relation has been simplified to a matrix multiplication form $p = Pf$ in the context of CBCT reconstruction, where $P$ is a projection matrix. However, in reality, the mapping is highly complicated due to numerous physical processes involved in the data acquisition process, e.g. scatter, beam hardening, detector response, etc. Hence, it is quite difficult to invert the mapping, i.e. retrieve the volumetric image $f$ based on projection images. Especially when there is only one projection image available, it is impossible to reconstruct the image from conventional CBCT reconstruction point of view.

The fundamental idea of our approach is to learn, via a sparse learning technique, an inverse mapping $f = P^{-1}[g] \equiv M[p]$. Specifically, the following aspects are included in our model. 1) We would go through the deformation vector domain, where $f(x) = f_0(x + v)$. Here $v$ is a vector field that deforms a prior image $f_0$. This approach allows us to utilize the PCA-based lung motion model (Söhn *et al.*, 2005; Zhang *et al.*, 2007; Li *et al.*, 2011c), under which $v$ is linearly related to the PCA coefficients $w$ and the degrees of freedom is substantially reduced, as it has been demonstrated that only a few PCA components are sufficient to describe the lung motion to a satisfactory level of accuracy. 2) The mapping between a set of PCA coefficients $w$ and a projection image $p$ is assumed to be linear, $w = Sp$, which is determined through a sparse learning process using prior 4D-CT data of the patient.

In the rest of this section, we will first briefly review the PCA representation of the lung motion model. We will then present our method to build the mapping. To demonstrate the feasibility of this approach, we focus our study on a scenario that all the projections are taken at a fixed projection angle. The potential application of this approach could be imaging the lung area during the delivery of Intensity Modulate Radiation Therapy (IMRT), where the linear accelerator gantry is fixed at one angle during delivery of a beam, so are the x-ray source and the detector.

*2.1 PCA based lung motion model*

Suppose there are $N$ phases of 4D-CT images in one breath cycle. The DVFs between the $N$ phases and a reference phase (e.g., the end of exhale phase) can be obtained by deformable image registration using, e.g., Demons algorithm (Gu *et al.*, 2010). The resulting DVFs for these phases form a matrix $V$, where each column represents the whole DVF field for a corresponding phase. Conducting PCA on this matrix yields:

$$v_n \approx \bar{v} + \sum_{k=1}^{K} u_k w_{k,n}\, , n = 1,2, \ldots\ldots, N, \qquad (1)$$

where $u$ and $w$ represent PCA eigenvectors and coefficients, respectively. The subscript $k$ is the index for the principal component, and $K$ is the number of principal components that one selects to approximate the motion vector field. Throwing away components with large $k$ values also removes noises and inaccuracy in the vector field that may be caused by registration error. According to previous studies (Zhang *et al.*, 2007; Li *et al.*, 2011c), $K = 2\sim3$ is adequate to accurately describe respiratory motion. We have therefore used $K = 3$ throughout our work. In a matrix form, Eq. (1) can be written as,

$$v = \begin{bmatrix} A & \bar{v} \end{bmatrix} \begin{bmatrix} w \\ 1 \end{bmatrix}, \qquad (2)$$

where $A$ is a matrix consisting of principal components.

*2.2 Projection partitioning*

To generate projection data used to build the model, we compute an x-ray projection for each 4D-CT image at the given beam angle using Siddon's ray-tracing algorithm (Siddon, 1985). We would like to maximally extract information from these projections and correlate it with the motion vector field. As such, we propose to use patches on the projection image to predict motion, as opposed to using the entire image. The rationale behind this strategy comes from the following observation: the projection contains both areas with large motion, such as those around diaphragm and chest-wall, and areas with small or no motion, such as air outside the patient body. The latter offer no extra motion information and may even 'dilute' the motion information provided by those areas with strong motion information, reducing prediction accuracy and robustness. Hence, we would like to use only a few small rectangular areas for motion prediction purposes. As such, we partition the projection image with two sets of interlaced rectangular grids as illustrated in Figure 1. Suppose the projection image is of a resolution $U\times V$ pixels, we partition it into small patches, each of a size $u\times v$ pixels. In order to capture the motion information, we partition the projection twice with two grids, one start at a location [1, 1] and the other start at [$u$/2, $v$/2]. Note that there are many other ways to partition the projection image. However, for simplicity we only use two grids mentioned here, and these two interlaced grids are found to be sufficient in our studies.

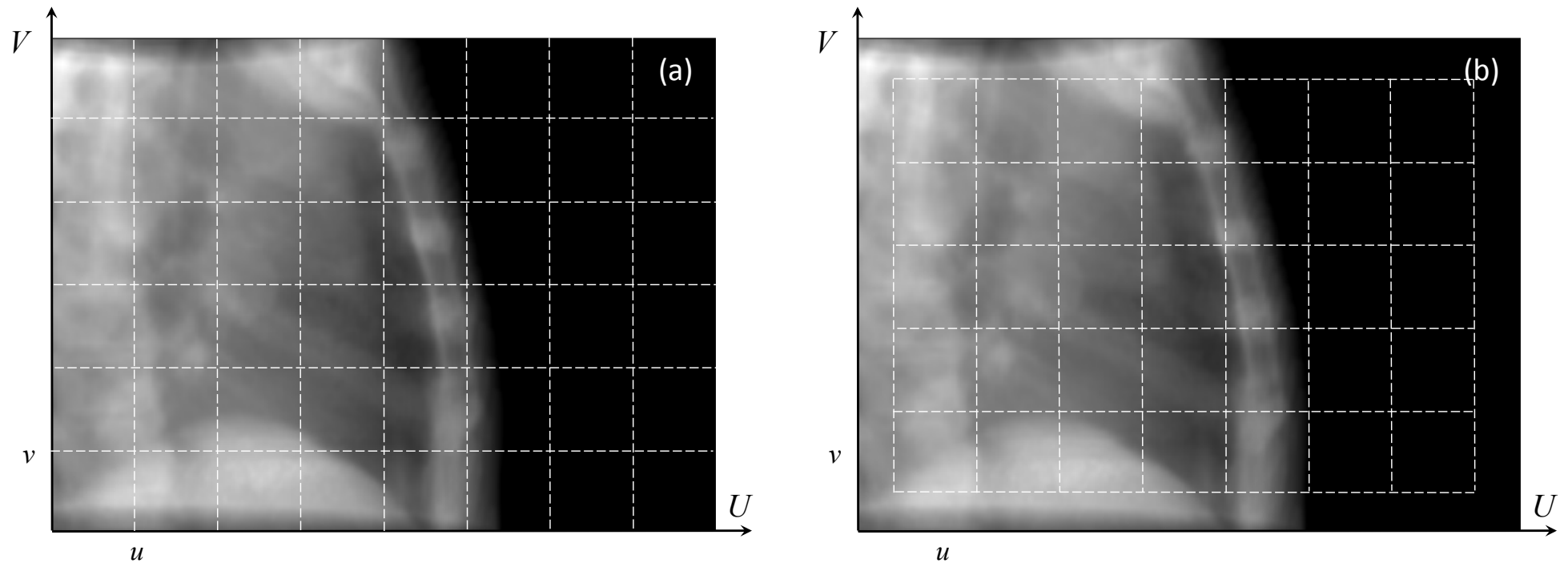

**Figure 1.** Illustration of how a projection image is partitioned into small rectangular patches by two grids shown in (a) and (b).

### *2.3 Sparse learning model*

As mentioned previously, different patch has different degrees of correlation with the PCA coefficients of the DVFs. There are two issues one needs to consider when building the correlation model in this training stage: 1) select those patches that are useful in terms of predicting the motion, and 2) build the motion model. We would like to address these two issues simultaneously via a sparse learning approach.

For the projection image at a given phase $n$, we extract a long column vector $p^n$ of length $J \equiv H\times M$, where $H = u\times v$ is the number of pixels per patch, and $M$ is the number of patches considered. The vector $p^n$ consists of $M$ subvectors. Each of them is of length $H$ and corresponds to the $H$ intensity values in a patch. These column vectors $p^n$ form a matrix $P \coloneqq [p^1; p^2; \dots; p^N] \in \mathbb{R}^{J\times N}$. Meanwhile, we also use PCA coefficients $w_{k,n}$ of the DVFs to form a matrix $W \in \mathbb{R}^{K\times N}$, where each column of $W$ is for one breathing phase $n$. Our assumption of a linear relationship between the projection image and the PCA coefficients translates to a linear equation $XP = W$, with $X \in \mathbb{R}^{K\times J}$. Moreover, the matrix $X$ can be partitioned into $M$ submatrices $X = [X_1; X_2; \dots; X_M]$, and $X_m \in \mathbb{R}^{K\times H}$ is for a patch $m$. If there is a non-zero entry in the submatrix $X_m$, it implies that the patch corresponding to this submatrix contributes to the motion model.

Under this structure, the assumption that only a few patches are important to predict the motion is translated to that only a few sub-matrices of $X$ is non-zero. It is hence our purpose to find such an $X$ matrix with this desired structure, such that $XP{\sim}W$. As such, we define an objective function $\sum_{m=1}^{M} \|X_m\|_F$, where $\|.\|_F$ denotes matrix Frobenius norm. It has been demonstrated previously that minimizing the objective function of this form would encourage the sparseness of the solution matrix among different sub-matrices, whereas the sparse requirement within each sub-matrix is not imposed. The applications of a norm defined as such have been recently explored in many problems, such as beam orientation optimization for IMRT (Jia *et al.*, 2011) and surface marker locations selection for tumor motion estimation (Dong *et al.*, 2012).

With this objective function defined, we formulate our problem as:

$$X^* = \arg\min_X \sum_{m=1}^{M} \|X_m\|_2 , \quad s.t.\, XP = W , \tag{3}$$

In reality, since the training data always contains noise, we solve an unconstrained optimization problem instead:

$$X^* = \arg\min_X \frac{1}{2} \|XP - W\|_2^2 + \lambda \sum_{m=1}^{M} \|X_m\|_2 , \tag{4}$$

to allow deviations of the predicted PCA coefficients $X^*P$ from the training data $W$. The positive constant $\lambda$ is manually selected to balance the fidelity term $\frac{1}{2}\|XP-W\|_2^2$ and the sparse term $\sum_{m=1}^{M}\|X_m\|_2$. To solve this optimization problem, we use the "glLeastR" algorithm in the package developed by Liu et al (Liu *et al.*, 2009).

*2.4 Measured projection preprocess*

Before using the trained model to predict a volumetric image based on one projection image, there is one more problem to solve. The projection images in the aforementioned training stage are generated by ray-tracing algorithm and are free of scatter. Yet, the projection acquired based on which a volumetric image is to be determined comes from measurements directly. There is an intensity inconsistency between the measured projections and those in the training stage. A baseline correction step has to be performed to correct the intensity of the measured projections.

For this purpose, we have developed the following strategy. We first acquire a number of projections in over one breath cycle and average these projection images. We then partition the averaged projections into patches in the same way as presented in Section 2.2. Meanwhile, we average the projection images used in the training stage. These two average images differ by a background. We would like to estimate a linear transform for each patch such that, when it is applied to the patch in the measured projection, the resulting patch and the corresponding patch in the training stage have the same mean value and standard deviation. Ideally, we would like to match the histogram of the two patches. But practically it is found that only matching the first two statistical moments, i.e. mean value and standard deviation, yield satisfactory background correction. Specifically, for a patch $m$ with $q_m$ denoting the vector containing its intensity values, the transformed patch becomes $p_m$:

$$p_m = a(m)q_m + b(m)\,, \tag{5}$$

where $a(m)$ and $b(m)$ are the parameters for this patch to be estimated. Using the patch intensity vector from the averaged training projection $\bar{p}^m$ and that from the measured projections $\bar{q}^m$, it can be derived (Zhen *et al.*, 2012) that:

$$a(m) = STD(\bar{p}^m)/STD(\bar{q}^m), \tag{6}$$

$$b(m) = E(\bar{p}^m) - a(m)E(\bar{q}^m). \tag{7}$$

here $E(.)$ and $STD(.)$ denote the calculations of mean value and standard deviation of a vector, respectively.

*2.5 Generate a volumetric image with one measured projection*

With the model trained and the patch intensity baseline correction parameters, we are ready to generate a volumetric image based on one measured projection image. Specifically, once a projection image is acquired, the patches that are relevant to the motion prediction are selected. Depending on their locations, corresponding intensity correction parameters are used to adjust the patch intensities. The resulting patches are then multiplied with the trained $X^*$ matrix, yielding the corresponding PCA coefficients of the deformation vector field. A vector field is generated, which is finally applied to the reference image, yielding the final volumetric image corresponding to the projection.

### *2.6 Evaluations*

#### *2.6.1 Synthetic study*

To evaluate our algorithm, we have generated a test dataset based on 4D-CT data of a real patient case. Specifically, the 4D-CT images are used in the training stage as presented previously. To test our algorithm, volumetric images at a much higher temporal resolution is needed. As such, we first compute ten DVFs between a reference phase (end of exhale) and each of the ten phases of the 4D-CT using demons deformable image registration algorithm. Then we interpolated five new DVFs between two adjacent phases by cubic spline interpolating. The total 60 DVFs are applied to the reference image, yielding 60 volumetric images corresponding to the variations of patient anatomy during a breathing cycle.

Corresponding to the 60 images in a breathing cycle, we synthesize x-ray projections using an in-house developed software tool (gDRR) (Jia *et al.*, 2012b) under the geometry of Varian TrueBeam On-Board-Imaging system (Varian Medical System Inc. Palo Alto, CA). In gDRR package, the projections are calculated using Monte Carlo method with various features in a realistic scan considered, such as energy spectrum, source fluence map, detector response, as well as scatter and quantum noise. The x-ray images have a resolution of $512\times384$ pixels with a physical size of $40\times30\ \text{cm}^2$. These simulated projections are then used to generate corresponding volumetric images using our algorithm.

For this synthetic data set, we have the ground truth about the PCA coefficients of the DVFs, as well as the images. These will be used to evaluate the results generated by our algorithm.

#### *2.6.2 Experimental study*

We have also tested our algorithm on a moving lung phantom. The phantom is constructed with an air cavity and materials equivalent to natural bone and soft tissues. A cylindrical tumor with ~3 cm in diameter by ~3 cm in height is attached to the end of a Styrofoam rod, which is put inside the chest cavity and programmed to move in three dimensions under the control of three stepper motors. In the experiment, the programmed trajectories of the tumor are all sinusoidal waves in three directions, with amplitudes in SI, RL and AP of 20mm, 5mm, and 5mm, respectively.

The 4D-CT dataset of the physical phantom was acquired using a Philips Brilliance 16 CT Scanner. The x-ray projections were acquired using Varian TrueBeam On-Board-Imaging system. The x-ray imager has a resolution of $1024\times768$ pixels with a physical size of $40\times30\ \text{cm}^2$. The projections are downsampled to $512\times384$ pixels. We first collect 659 projections in half-fan mode with 125 kVp, 60 mA, and 20 ms exposure time and a CBCT image is reconstructed using FDK algorithm (Feldkamp *et al.*, 1984). The purpose of this CBCT image is to allow for a rigid registration with the averaged 4D-CT images. After that, projections were acquired in Dynamic Gain Fluoro mode with 125 kVp, 60 mA, and 20 ms exposure time at a frequency of 11 frames per second. The projection angle is at 0 degree. The projections at the first 5 sec are used for intensity baseline corrections, and the rest are used to test our method.

## 3. Results

### *3.1* Synthetic *patient case*

#### *3.1.1 Sparse learning results*

Figure 2 shows the results of sparse learning. We use a partition indicated in Figure 2 with a square size of $4.97{\times}4.97\ \mathrm{cm}^2$. The reason for this choice will be presented in Section 3.3.2. After running the sparse learning algorithm using the ten 4D-CT images in a breathing cycle, it was found that the patches indicated by the black rectangle are automatically selected and a matrix $X^*$ is determined at the same time to map their intensities into the PCA coefficients of the DVFs. From this figure, we can see that our algorithm selects the patches that probably contain the richest motion information among all the patches, namely those around the diaphragm.  This result reflects the efficacy of our algorithm to a certain extent.

#### *3.1.2 Motion prediction results: PCA coefficients*

After sparse learning, we can use the built model to predict the motion for each input projection image and hence generate a corresponding volumetric image. We first show in Figure 3 (a)~(c) the 1$^{st}$ , 2$^{nd}$, and 3$^{rd}$ calculated PCA coefficients at 60 time points within a breathing cycle using the corresponding projections. In this synthetic case, the ground truth values are available, and hence we also present in Figure 3(d), (e), and (f) residual errors. Quantitatively, Table 1 presents the errors in different cases, where the error is defined as $e = \|w - w_0\|/\|w_0\|$. Here $w$ is a vector consisting of the PCA coefficients of interest, and $w_0$ is the ground truth one. $\|.\|$ denotes standard L2 norm. Based on the figures and the table, we have the following observations. 1) Comparing the results with and without patch-based intensity correction, it is found that there is a systematic error in the prediction results if the intensity is not corrected. This is because that in the linear motion model, a background change in the projection images result in a systematic deviation of the predicted results. Similar results have also been observed in other experiments, which demonstrate the necessity of performing a background match step. Hence, in the rest of this paper, we will not show the results without background matching, and will focus on the sparse learning only. 2) Comparing the results with and without sparse learning, both after the background intensity correction, it was observed that sparse learning model is able to reduce the prediction error of the 1st PCA coefficients. Although the error in the second and the third PCA components are slightly bigger, since the motion is dominated by the first component, the overall results in DVF will be better, as will be shown later.

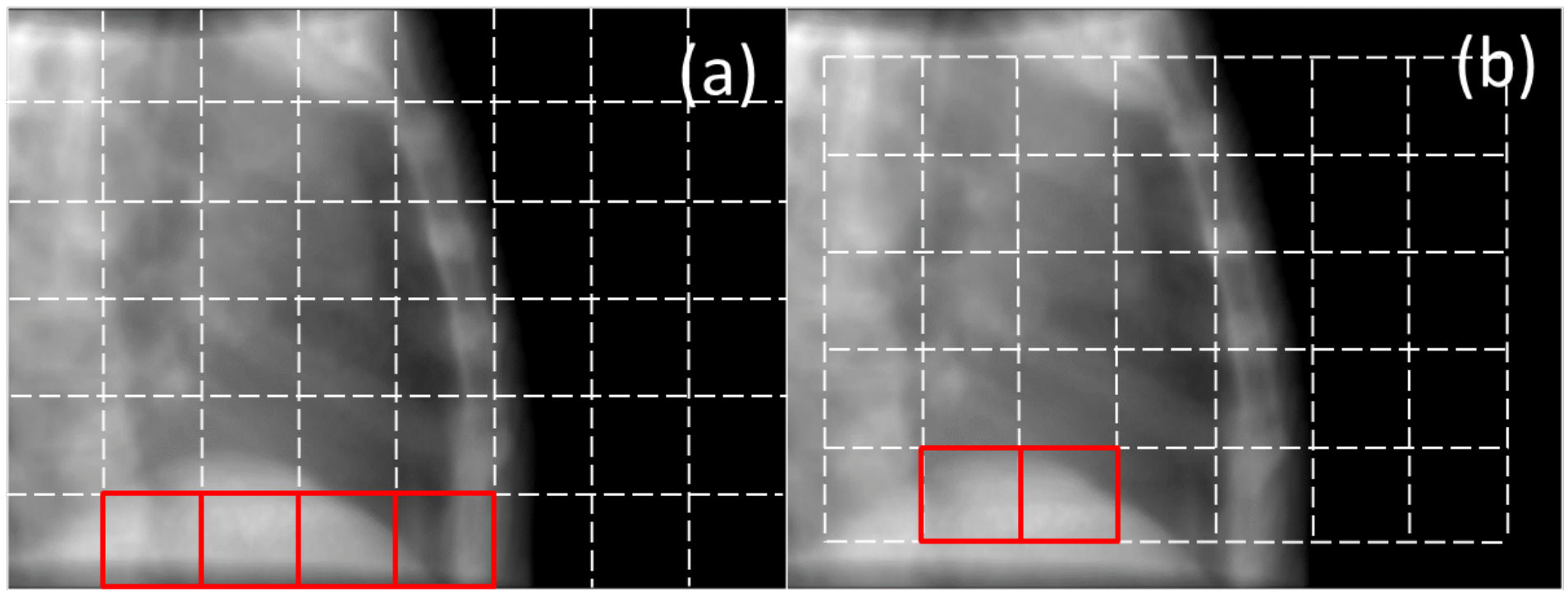

**Figure 2.** Patch selection result. In (a) and (b), the patches indicated by the rectangles are selected by sparse learning algorithm.

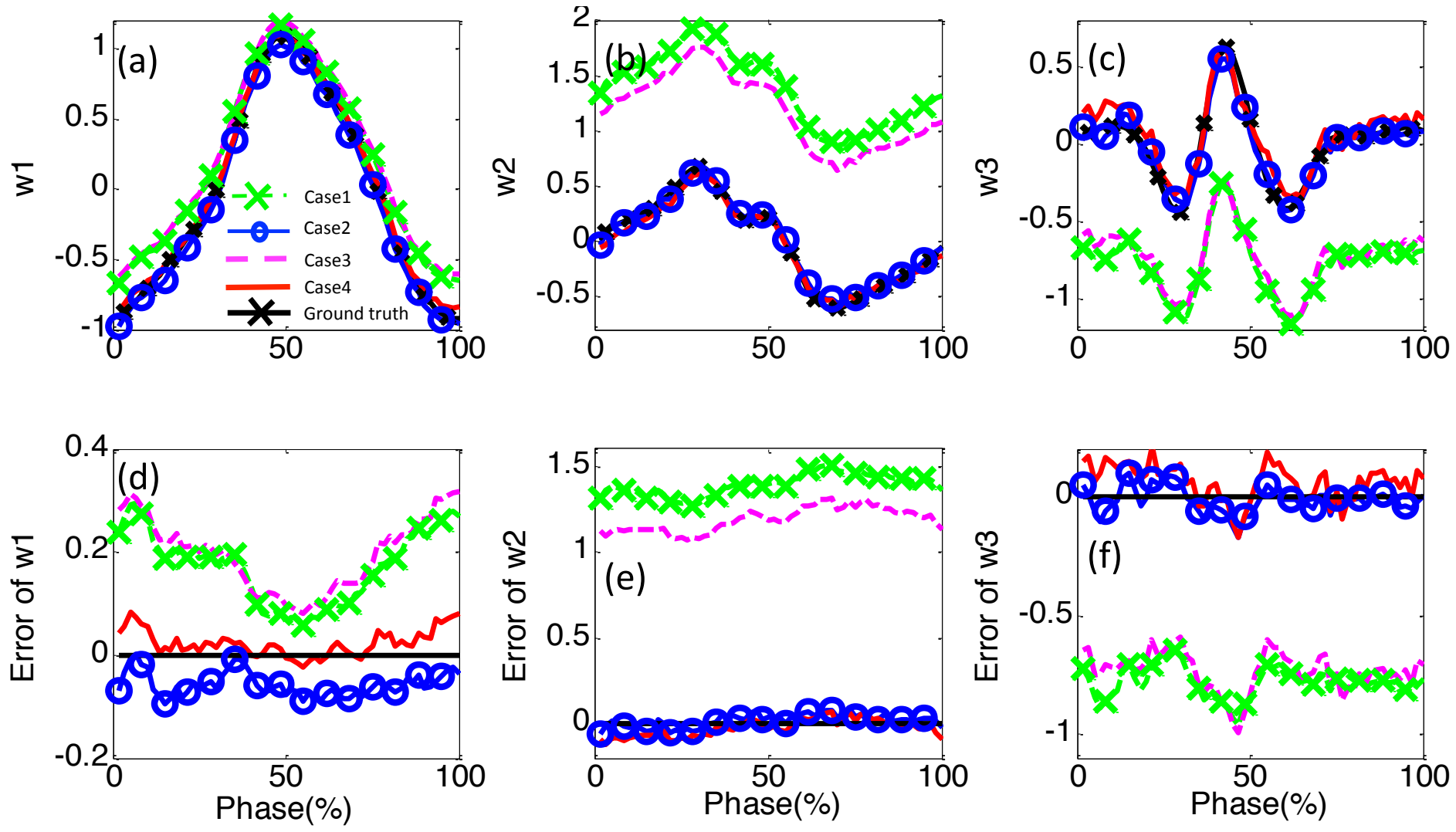

**Figure 3.** (a)~(c) The 1st , 2nd , 3rd PCA coefficient results. (d)~(f) The residual error of the 1st, 2nd, 3rd PCA coefficient results comparing with the ground truth.

**Table 1.** Errors in the predicted PCA coefficients in different cases.

| Cases | | | PCA coefficient error | | |
|---|---|---|---|---|---|
| Number | Sparse training | Intensity correction | First | Second | Third |
| 1 | Yes | No | 0.27 | 3.55 | 3.08 |
| 2 | No | Yes | 0.10 | 0.11 | 0.22 |
| 3 | No | No | 0.31 | 3.03 | 2.89 |
| 4 | Yes | Yes | 0.05 | 0.14 | 0.38 |

### *3.1.2 Motion prediction results: Images*

The 60 volumetric images can be accordingly generated using the predicted motion vector fields. We compare these images with the corresponding ground truth ones. Since each test image is generated by deforming the reference image and the error in the predicted DVF is inevitable, it is expected that there will be structural location inaccuracy. We quantify the comparison using a generalized gamma index (Low *et al.*, 1998) to this context, which considers differences between two images with both intensity difference and spatial difference considered. Specifically,

$$\gamma(x) = \min_y \sqrt{\left[\frac{f_0(x)-f(y)}{IC}\right]^2 + \left[\frac{x-y}{DC}\right]^2}, \quad (5)$$

where $f$ and $f_0$ are the test and the ground truth images. $IC$ and $DC$ are intensity and deformation criteria for the test, which are chosen as 10HU and 0.5mm, respectively. It is understood that a voxel at $x$ passes the test, if $\gamma(x) < 1$. We report the passing rate of the whole image, i.e. the percentage of the voxels that pass the test.

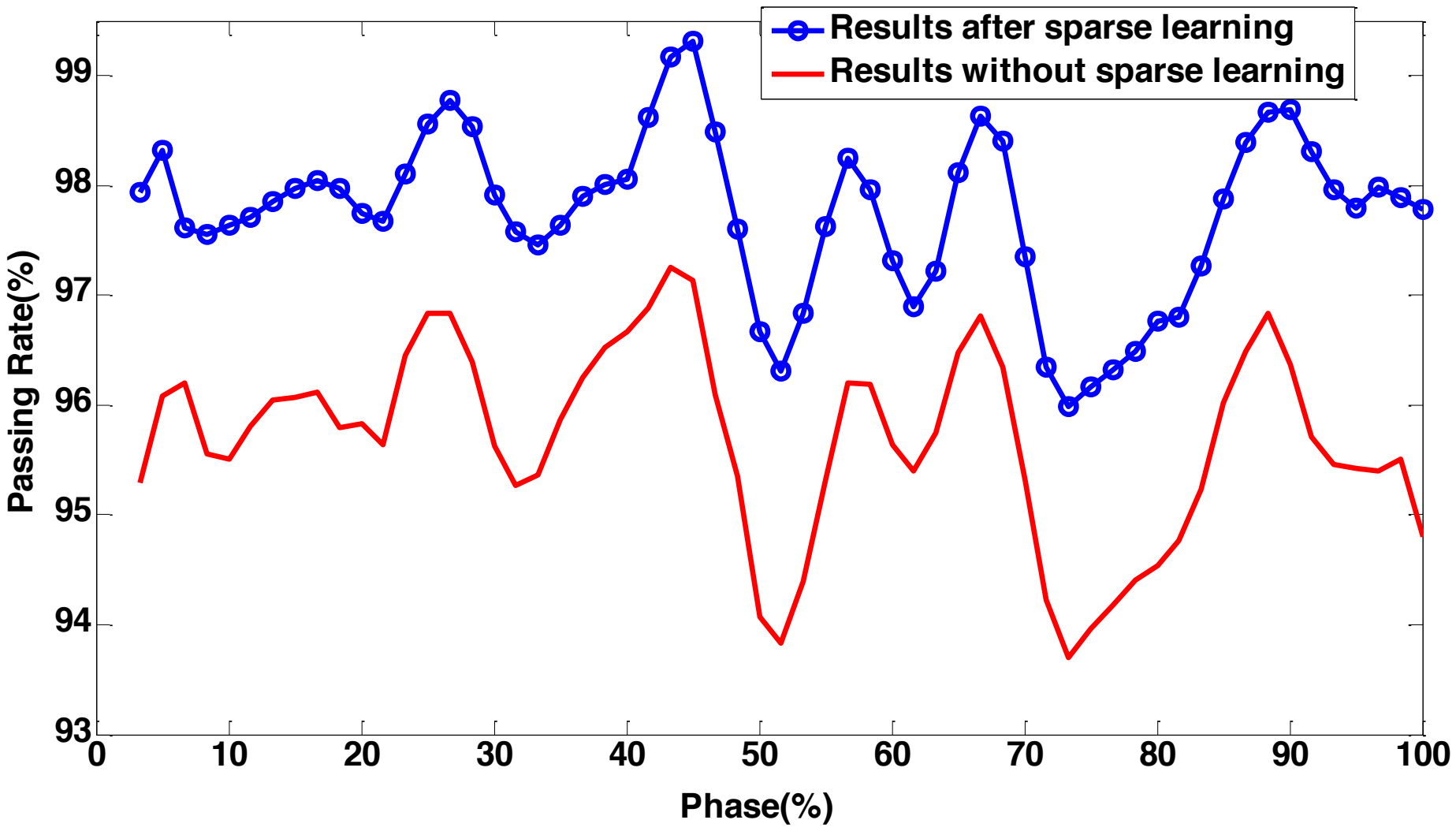

**Figure 4.** The gamma index result of predicted images in one breath cycle.

**Table 2.** Errors of each voxel in vector field. (Result with sparse learning/Result without sparse learning).

| | $e_{max}$ (mm) | $\bar{e}$ (mm) | $e_{95\%}$ (mm) |
|---|---|---|---|
| SI | 2.12/3.47 | 0.08/0.14 | 0.64/1.80 |
| RL | 0.87/1.22 | 0.03/0.08 | 0.37/1.14 |
| AP | 1.07/1.47 | 0.05/0.13 | 0.45/1.21 |
| Amplitude | 2.22/3.74 | 0.09/0.23 | 0.92/2.40 |

Figure 4 shows the gamma index results of the predicted images in one breath cycle. It is clear that the proposed sparse learning model outperforms the other one, in that the gamma test passing rate is consistently higher. We further examined the improvement in vector field result quantitatively. Maximum error $e_{max}$, 95$^{\text{th}}$ percentile error $e_{95\%}$, and mean error $\bar{e}$ are reported in Table 2 for the SI, RL, AP component of the vector field, as well as for the amplitude. It is observed that the sparse learning method can greatly reduce the errors at all items.

### *3.2 Real phantom case*

For the real phantom case, the ground truth about the PCA coefficients, DVFs and the images are not available. However, we have the knowledge about the tumor motion trajectory, as it was controlled by the step motors to move with sinusoidal curves in three directions. To derive the motion trajectory in the reconstructed images, we segment the tumor in the reconstructed images using a simple threshold method, and the centroid of the tumor is calculated. In Figure 5(a), the 3D tumor trajectories obtained in the methods with and without using the sparse learning method are compared with the ground truth. It is found that the one with the sparse learning method yields a trajectory closer to the ground truth. In addition, we have also plotted the motion coordinates along the SI, RL and AP directions as functions of breathing phase in Figure 5(b)~(d). Again, the sparse learning method yields better agreements with the ground truth. To further quantify the improvements, maximum error, 95-percentile error, and mean error along the SI, RL, AP direction, as well as the magnitude are summarized in **Table 3**. It is observed that the sparse learning method can greatly reduce the errors and hence increase tumor location accuracy. Especially the mean error by our proposed method can be reduced to 0.82 mm, comparing to the 1.66 mm error without using sparse learning.

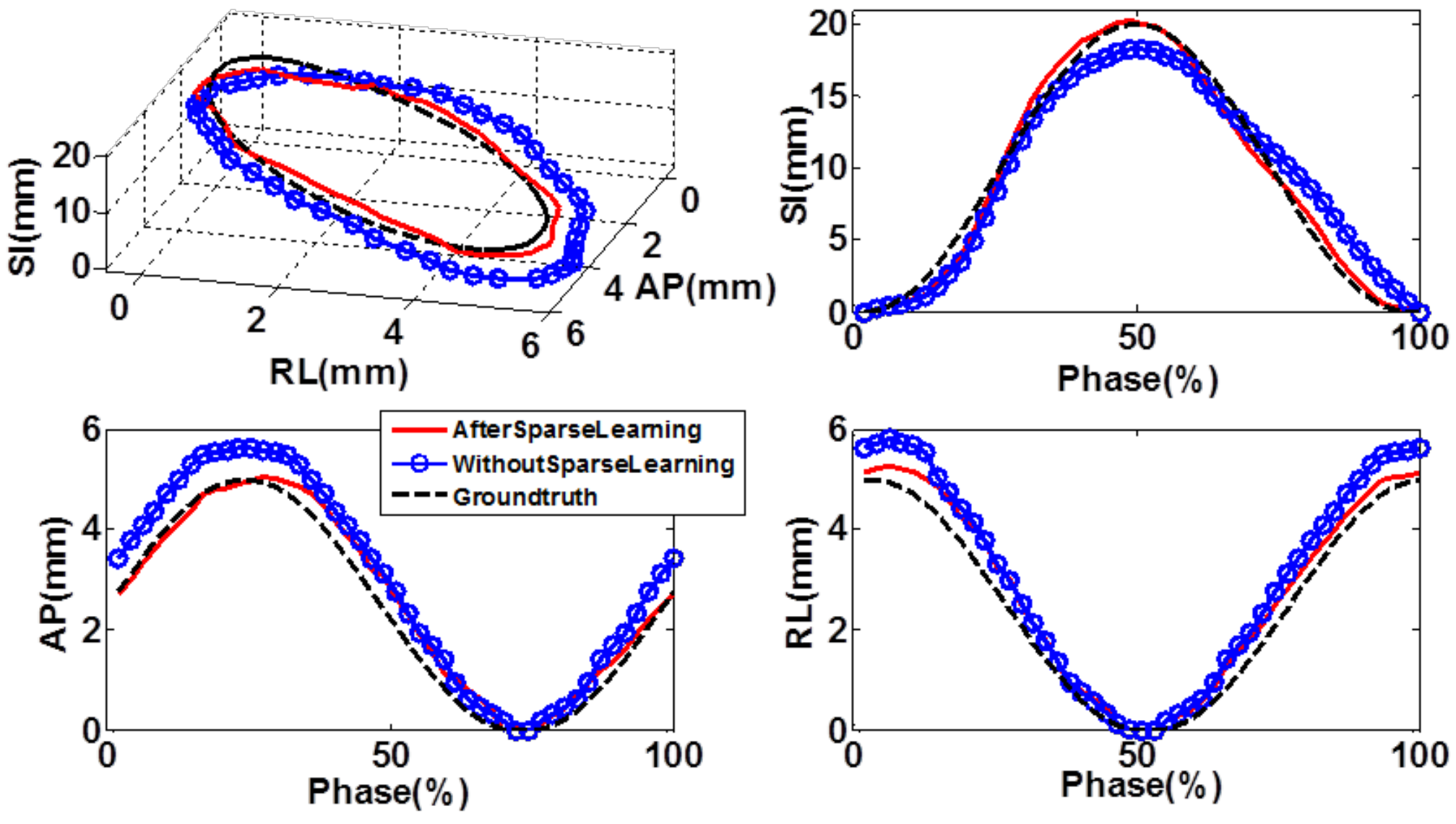

**Figure 5.** The tumor center's trajectory, and SI, RL, AP direction, respectively.

**Table 3**. Errors in tumor localization in the real phantom case. (Result after sparse learning/Result without sparse learning)

| | $e_{max}$ (mm) | $\bar{e}$ (mm) | $e_{95\%}$ (mm) |
|---|---|---|---|
| SI | 1.69 /3.26 | 0.66 /1.39 | 1.42/3.11 |
| RL | 0.67/1.01 | 0.31/0.51 | 0.61/0.89 |
| AP | 0.69/0.79 | 0.27/0.54 | 0.56/0.77 |
| Amplitude | 1.74/3.32 | 0.82/1.66 | 1.55/3.14 |

*3.3 Robust analysis*

In this section, we would like to examine how the proposed sparse learning method is robust against different parameters.

*3.3.1 Effects of sparse level*

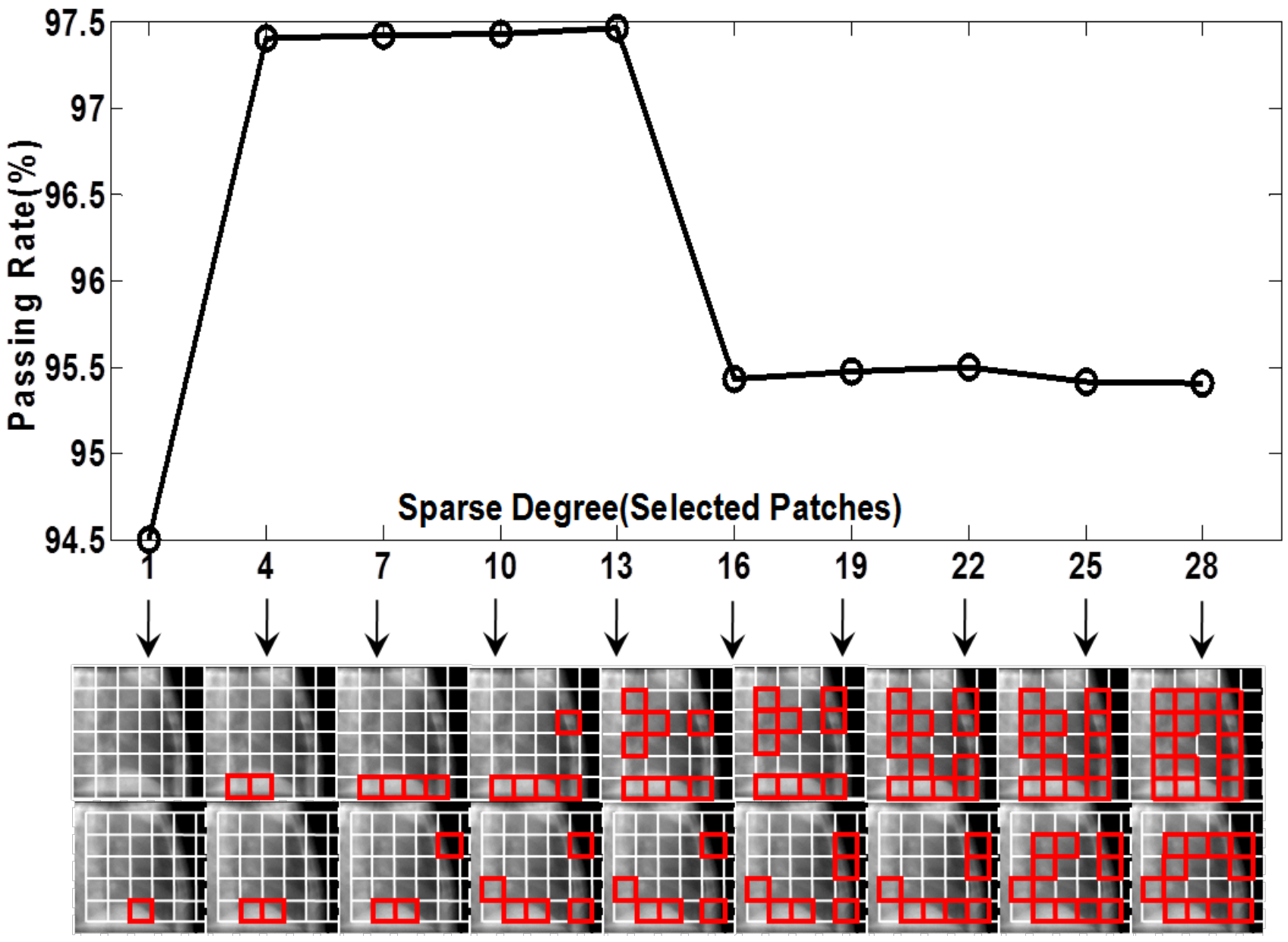

**Figure 6.** Gamma index result as the function of patches number selected by sparse learning. The patches indicated by the rectangles are selected by sparse learning algorithm.

It is necessary to balance the fidelity term and the sparse term in Eq. (4) by adjusting the parameter $\lambda$. Hence, it is interesting to see how this weighting factor affects the results. By tuning the parameter, the number of selected patches varies. As shown in Figure 6, we have calculated the averaged gamma index passing rate at different sparse degree. Correspondingly the selected patches are also plotted. The results show that the high passing rate above 97% can be achieved using 4 to 13 patches. When the algorithm emphasizes too much on the sparseness such that only one patch is selected, the patch does not contain enough motion information and hence the passing rate is low. In contrast, when a lot of patches are selected, the model approaches the one without sparse learning, yielding again a low level of passing rate. It is interesting to observe that, regardless the number of patches selected, there are some key locations that are always chosen by the algorithm, e.g. those around the diaphragm area. This also demonstrates the correctness of our patch selection algorithm to a certain extent

*3.3.2 Effects of patch size*

We have also studied the impacts of patch sizes. For each given combination $u, v$ in the unit of pixel number, we conduct the training process and then use the trained model to generate a set of images. The gamma passing rate under each $u, v$ is shown in gray scale in Figure 7. It is observed that there exists an optimal patch size of 64×64 pixel$^2$ under which the highest passing rate is achieved. This is the size we used in previous studies. When the patch is too small, the selected patches do not contain sufficient motion information, as the key features that reflect motion, e.g. edges, sometimes move out of the selected patch. In contrast, a large patch with too many pixels dilutes the motion information contained in it. In either case, the passing rate is relatively low.

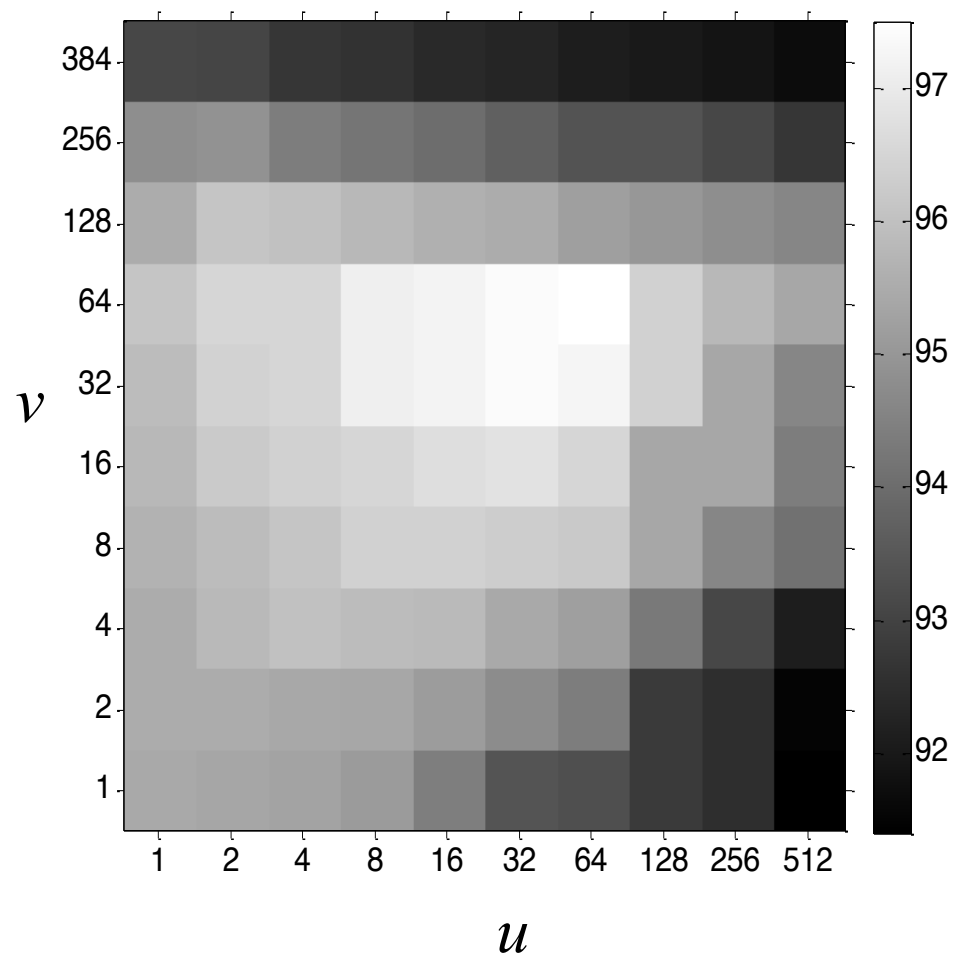

**Figure 7.** The gamma index result as a function of patches sizes.

*3.3.3 Impacts of diaphragm*

Although it has been demonstrated that the patch selection algorithm can automatically choose patches with motion information, e.g. in Figure 2 and Figure 6, diaphragm has been a very clear motion feature in the projection images, which makes the selection easy in those cases. We would like to examine if the algorithm is robust enough in the absence of the diaphragm. As such, we mask out the entire diaphragm area in the projections, and conduct the study again. The results are in Figure 8. It is found that the patches selected by sparse learning are divided in to two parts, in which the first three patches are in the wall chest area, and the other two are in the area near the diaphragm. Figure 9 shows the gamma index results of the predicted images in one breath cycle. The result shows that the proposed sparse learning model outperforms again the one without sparse learning. Yet, excluding the diaphragm area definitely makes the problem harder, which is clearly indicated by the reduced passing rates in Figure 9 compared with those in Figure 4 for both cases. Figure 9 also shows that the improvements of sparse learning are more obvious in the absence of diaphragm. Under such a context, the sparse learning model can still identify patches with important motion information and only use them to predict lung motion. In contrast, the one without sparse learning uses all the patches. Since motion information is weak, they are easily impeded by other irrelevant information, e.g. noise.

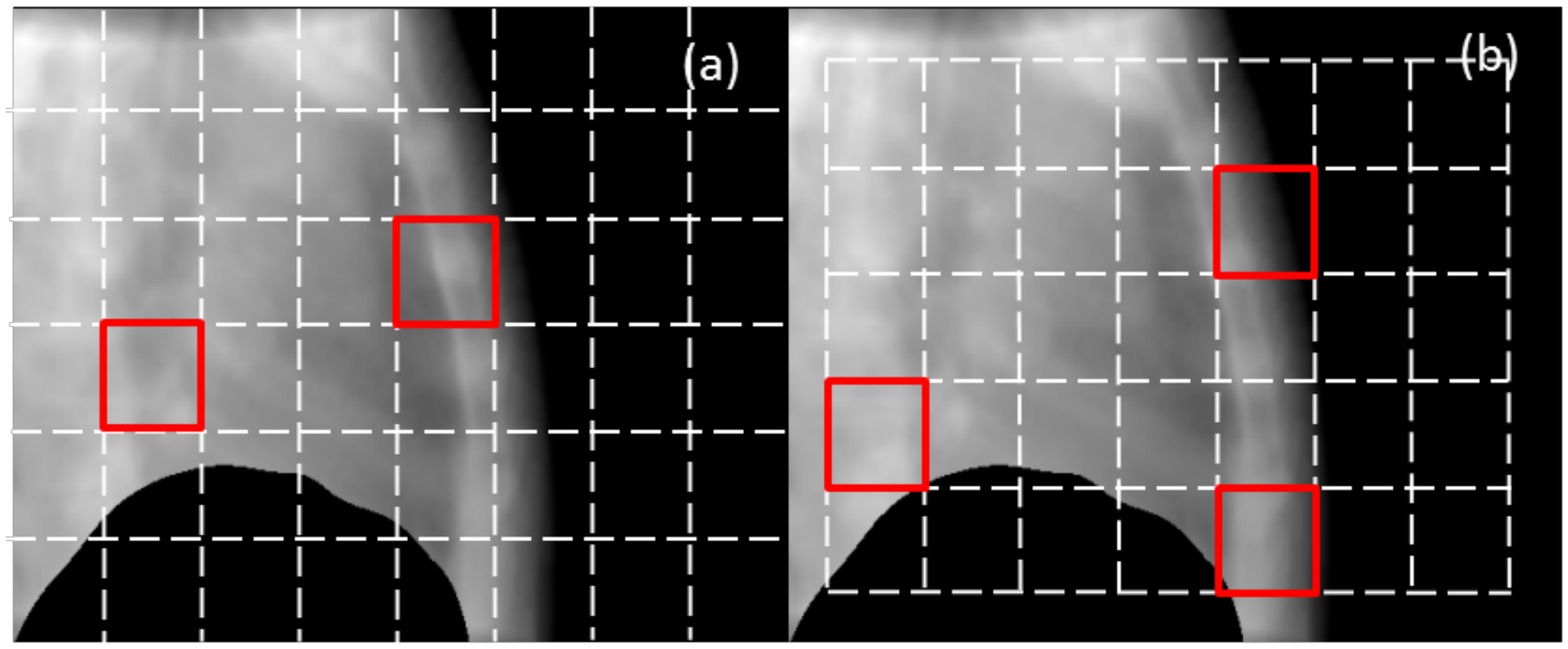

**Figure 8.** Patch selection result when removing diaphragm. In (a) and (b), the patches indicated by the black rectangles are selected by sparse learning algorithm.

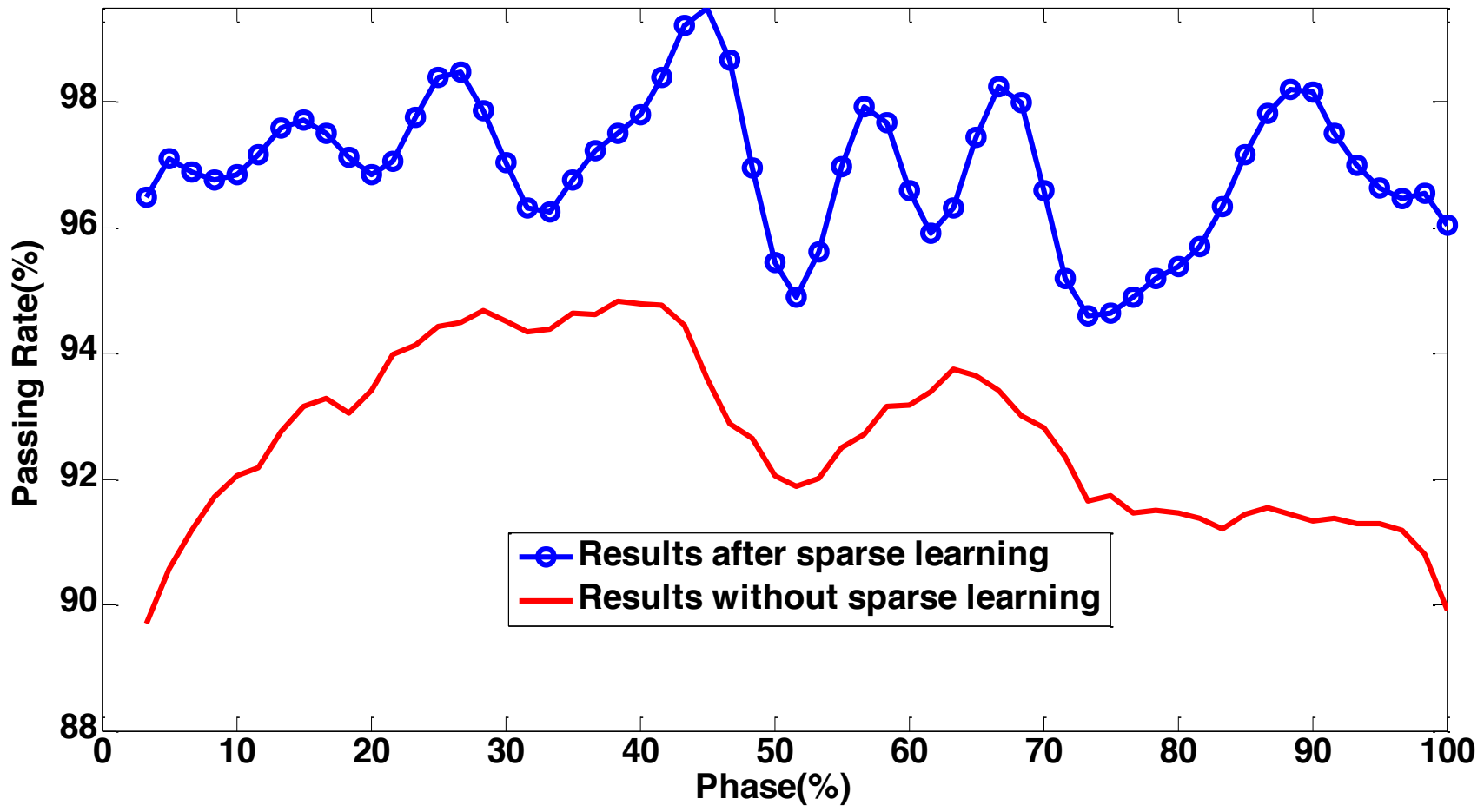

**Figure 9.** The gamma index result of prediction phase image in one breath cycle when in the absence of the diaphragm.

*3.4 Computation time*

Finally, we report the computation time for our algorithm in both the simulation and the experimental cases. Specifically, the time spent on training the model is broken down to five parts for (1) deformable image registration, (2) PCA of the vector field, (3) simulate projections via DRR, (4) sparse learning, and (5) intensity correction, respectively. Once the model is built, it is applied to a projection to yield a volumetric image. The time on this step (6) is also reported. Among them, steps (1), (3) are achieved on an Nvidia Titan GPU card using our previously developed packages (Gu *et al.*, 2010) (Jia *et al.*, 2012b). Step (6) is also performed on GPU as it contains simply matrix-vector operations and volume interpolations, which are GPU friendly. Other steps are conducted on a CPU platform in MATLAB. Detailed computation time is shown in Table 4. The time spent on the model training part is relatively long due to many complicated computations. Especially, the deformable image registration step dominates the calculation time, as 10 volumetric images are registered. PCA and sparse learning is executed in MATLAB, which can be further sped up with optimal implementations. The DRR part is also long, as polyenergetic projections are computed in our study. It will be our future research to accelerate computations in this model training part. In contrast, once a model is built, it takes only 0.09 sec to predict a volumetric image on GPU based on a projection image, which implies the feasibility of using our model in online applications.

**Table 4**. Computation time of our method.

| Case | Model training (sec) | | | | | Generate a volumetric image (sec) |
|---|---|---|---|---|---|---|
| | Registration | PCA | DRR | Sparse learning | Intensity correction | |
| Simulation | 78.3 | 5.28 | 8.26 | 18.9 | 0.14 | 0.09 |
| Experiment | 75.5 | 5.15 | 8.26 | 18.7 | 0.14 | 0.09 |

## 4. Conclusions

In this paper, we have presented a method aiming at solving two problems in our previously-developed algorithm for single-projection based volumetric image reconstruction, namely robustness and computational efficiency. As opposed to obtaining a deformation vector field via a forward projection matching, a sparse learning approach is employed. To fully extract motion information hidden in projection images, we participate a projection into small patches. At a training stage, we utilize sparse learning to automatically select patches that have a high correlation with the PCA coefficients of the lung motion model. A motion model is built at this stage together with the patch selection. We have also proposed an intensity baseline correction method based on the partitioned projection, in which the first and the second moments of pixel intensities at a patch in a simulated image are matched with those in a measured image via a linear transformation. The proposed method has been validated in simulated data and real phantom data. It is found that the algorithm is able to identify patches that contain relevant motion information, e.g. diaphragm region. Intensity correction step is important to remove the systematic error in the motion prediction. After that, each intensity corrected projections can be used to generate a volumetric image. For the simulation case, the sparse learning model yields reduced prediction error in PCA coefficients and higher gamma passing rate in the reconstructed image compared to the model without sparse learning but with intensity correction. In the phantom case, the predicted tumor motion trajectory is successfully reconstructed with 0.82 mm mean error in 3D space. The algorithm robustness with respect to sparse level, patch size, and existence of diaphragm is also tested.

We would like to discuss a few future directions. First, patient studies are needed to. Considering the complexity of patient respiratory motion, e.g. amplitude/phase irregularities and baseline drift, only comprehensive tests in patient cases can evaluate the performance of the algorithm and hence its applicability in real clinical settings. Second, this current work only dealt with the case for a fixed projection angle. Its potential application could be to reconstruct volumetric images during IMRT beam delivery. When it comes to Volumetric Modulated Arc Therapy (VMAT), the beam rotates continuously around a patient. A model built for a particular projection angle will be invalid for other beam angles. Potential solution could be to build a number of models one for each projection angle. Yet, this brute force way certainly reduces the computational efficiency. It will also be our future research topic to develop novel methods to extent the current algorithm for the VMAT case.

**Acknowledgement**

This work is supported in part by NIH (1R01CA154747-01), National Natural Science Foundation of China (No. 81301940), and Guangdong Strategic emerging industry core technology research (No. 2011A081402003).

**References**

Ahmad M, Balter P and Pan T 2011 Four-dimensional volume-of-interest reconstruction for cone-beam computed tomography-guided radiation therapy *Medical Physics* **38** 5646

Cai J-F, Jia X, Gao H, Jiang S B, Shen Z and Zhao H 2012 Cine cone beam CT reconstruction using low-rank matrix factorization: algorithm and a proof-of-princple study *arXiv preprint arXiv:1204.3595*

Chen G-H, Tang J and Leng S 2008 Prior image constrained compressed sensing (PICCS): a method to accurately reconstruct dynamic CT images from highly undersampled projection data sets *Medical Physics* **35** 660

Chou C-R, Frederick B, Liu X, Mageras G, Chang S and Pizer S 2011 Claret: A fast deformable registration method applied to lung radiation therapy. In: *Fourth International (MICCAI) Workshop on Pulmonary Image Analysis,* pp 113-24

Chou C-R, Frederick B, Mageras G, Chang S and Pizer S 2013 2D/3D Image Registration using Regression Learning *Computer Vision and Image Understanding*

Dong B, Graves Y J, Jia X and Jiang S B 2012 Optimal surface marker locations for tumor motion estimation in lung cancer radiotherapy *Physics in Medicine and Biology* **57** 8201

Feldkamp L A, Davis L C and Kress J W 1984 Practical cone beam algorithm *Journal of the Optical Society of America A-Optics Image Science and Vision* **1** 612-9

Gao H, Li R, Lin Y and Xing L 2012 4D cone beam CT via spatiotemporal tensor framelet *Medical Physics* **39** 6943

Gu X, Pan H, Liang Y, Castillo R, Yang D, Choi D, Castillo E, Majumdar A, Guerrero T and Jiang S B 2010 Implementation and evaluation of various demons deformable image registration algorithms on a GPU *Physics in Medicine and Biology* **55** 207

Jaffray D A and Siewerdsen J H 2000 Cone-beam computed tomography with a flat-panel imager: Initial performance characterization *Medical Physics* **27** 1311-23

Jaffray D A, Siewerdsen J H, Wong J W and Martinez A A 2002 Flat-panel cone-beam computed tomography for image-guided radiation therapy *International Journal of Radiation Oncology Biology Physics* **53** 1337-49

Jia X, Men C, Lou Y and Jiang S B 2011 Beam orientation optimization for intensity modulated radiation therapy using adaptive l2, 1-minimization *Physics in Medicine and Biology* **56** 6205

Jia X, Tian Z, Lou Y, Sonke J-J and Jiang S B 2012a Four-dimensional cone beam CT reconstruction and enhancement using a temporal nonlocal means method *Medical Physics* **39** 5592

Jia X, Yan H, Cervino L, Folkerts M and Jiang S B 2012b A GPU Tool for Efficient, Accurate, and Realistic Simulation of Cone Beam CT Projections *Med. Phys.* **39** 7368-78

Jiang S B 2006 Radiotherapy of mobile tumors. In: *Seminars in radiation oncology*: Elsevier) pp 239-48

Keall P, Starkschall G, Shukla H e e, Forster K, Ortiz V, Stevens C, Vedam S, George R, Guerrero T and Mohan R 2004 Acquiring 4D thoracic CT scans using a multislice helical method *Physics in Medicine and Biology* **49** 2053

Keall P J, Mageras G S, Balter J M, Emery R S, Forster K M, Jiang S B, Kapatoes J M, Low D A, Murphy M J, Murray B R, Ramsey C R, Van Herk M B, Vedam S S, Wong J W and Yorke E 2006 The management of respiratory motion in radiation oncology report of AAPM Task Group 76 *Medical Physics* **33** 3874-900

Leng S, Zambelli J, Tolakanahalli R, Nett B, Munro P, Star-Lack J, Paliwal B and Chen G-H 2008 Streaking artifacts reduction in four-dimensional cone-beam computed tomography *Medical Physics* **35** 4649

Li R, Fahimian B P and Xing L 2011a A Bayesian approach to real-time 3D tumor localization via monoscopic x-ray imaging during treatment delivery *Medical Physics* **38** 4205

Li R, Lewis J H, Cervino L I and Jiang S B 2009 4D CT sorting based on patient internal anatomy *Physics in Medicine and Biology* **54** 4821

Li R, Lewis J H, Jia X, Gu X, Folkerts M, Men C, Song W Y and Jiang S B 2011b 3D tumor localization through real-time volumetric x-ray imaging for lung cancer radiotherapy *Medical Physics* **38** 2783

Li R, Lewis J H, Jia X, Zhao T, Liu W, Wuenschel S, Lamb J, Yang D, Low D A and Jiang S B 2011c On a PCA-based lung motion model *Physics in Medicine and Biology* **56** 6009

Li T, Koong A and Xing L 2007 Enhanced 4D cone-beam CT with inter-phase motion model *Medical Physics* **34** 3688

Liu J, Ji S and Ye J 2009 SLEP: Sparse Learning with Efficient Projections. (Arizona State University

Low D A, Harms W B, Mutic S and Purdy J A 1998 A technique for the quantitative evaluation of dose distributions *Medical Physics* **25** 656-61

Low D A, Parikh P J, Lu W, Dempsey J F, Wahab S H, Hubenschmidt J P, Nystrom M M, Handoko M and Bradley J D 2005 Novel breathing motion model for radiotherapy *International Journal of Radiation Oncology* Biology* Physics* **63** 921-9

McBain C A, Henry A M, Sykes J, Amer A, Marchant T, Moore C M, Davies J, Stratford J, McCarthy C and Porritt B 2006 X-ray volumetric imaging in image-guided radiotherapy: the new standard in on-treatment imaging *International Journal of Radiation Oncology* Biology* Physics* **64** 625-34

McClelland J, Hawkes D J, Schaeffter T and King A P 2012 Respiratory motion models: a review *Medical image analysis*

Pan T, Lee T-Y, Rietzel E and Chen G T 2004 4D-CT imaging of a volume influenced by respiratory motion on multi-slice CT *Medical Physics* **31** 333

Pan T, Sun X and Luo D 2007 Improvement of the cine-CT based 4D-CT imaging *Medical Physics* **34** 4499

Park J C, Park S H, Kim J H, Yoon S M, Kim S S, Kim J S, Liu Z, Watkins T and Song W Y 2011 Four-dimensional cone-beam computed tomography and digital tomosynthesis reconstructions using respiratory signals extracted from transcutaneously inserted metal markers for liver SBRT *Med Phys* **38** 1028-36

Ren L, Chetty I J, Zhang J, Jin J-Y, Wu Q J, Yan H, Brizel D M, Lee W R, Movsas B and Yin F-F 2012 Development and clinical evaluation of a three-dimensional cone-beam computed tomography estimation method using a deformation field map *International Journal of Radiation Oncology* Biology* Physics* **82** 1584-93

Rit S, Wolthaus J W, van Herk M and Sonke J-J 2009 On-the-fly motion-compensated cone-beam CT using an a priori model of the respiratory motion *Medical Physics* **36** 2283

Siddon R L 1985 Fast calculation of the exact radiological path for a 3-dimensional CT array *Medical Physics* **12** 252-5

Söhn M, Birkner M, Yan D and Alber M 2005 Modelling individual geometric variation based on dominant eigenmodes of organ deformation: implementation and evaluation *Physics in Medicine and Biology* **50** 5893

Staub D, Docef A, Brock R S, Vaman C and Murphy M J 2011 4D Cone-beam CT reconstruction using a motion model based on principal component analysis *Medical Physics* **38** 6697

Wang J and Gu X 2013 High-quality four-dimensional cone-beam CT by deforming prior images *Physics in Medicine and Biology* **58** 231

Yan H, Zhen X, Folkerts M, Li Y, Pan T, Cervino L, Jiang S B and Jia X 2014 A hybrid reconstruction algorithm for fast and accurate 4D cone-beam CT imaginga) *Medical Physics* **41** 071903

Zeng R, Fessler J A and Balter J M 2005 Respiratory motion estimation from slowly rotating X-ray projections: Theory and simulation *Medical Physics* **32** 984

Zhang Q, Pevsner A, Hertanto A, Hu Y-C, Rosenzweig K E, Ling C C and Mageras G S 2007 A patient-specific respiratory model of anatomical motion for radiation treatment planning *Medical Physics* **34** 4772

Zhang Y, Yang J, Zhang L, Balter P A and Dong L 2013 Modeling respiratory motion for reducing motion artifacts in 4D CT images *Medical Physics* **40** 041716

Zhen X, Gu X, Yan H, Zhou L, Jia X and Jiang S B 2012 CT to cone-beam CT deformable registration with simultaneous intensity correction *Physics in Medicine and Biology* **57** 6807-26

Zheng Z, Sun M, Pavkovich J and Star-Lack J 2011 Fast 4D cone-beam reconstruction using the McKinnon-Bates algorithm with truncation correction and nonlinear filtering. In: *Proc. SPIE,* p 79612U